\documentclass[aps,prd,superscriptaddress,nofootinbib,floats,preprint,floats,floatfix]{revtex4-1}

\usepackage{makecell} 
\usepackage{bm}
\usepackage{amsmath} 
\usepackage{amssymb}  
\usepackage{graphicx}
\usepackage{color}   
\usepackage{tensor}  
\usepackage{braket}  
\usepackage{multirow}
\usepackage{booktabs} 
\usepackage{tabularx}
\usepackage{subcaption}
\usepackage{caption}
\usepackage{float}
\usepackage{nameref}
\usepackage{hyperref}
\hypersetup{
	colorlinks=true,  
	linkcolor=red,    
	citecolor=blue,  
}

\allowdisplaybreaks[1]

\begin{document}
	
	\title{Gravitational waves and primordial black holes from the T-model inflation with Gauss-Bonnet correction}
	
	\author{Peng-Bo Chen}
	\email{cqyang1988@gmail.com}
	\affiliation{School of physics, Xidian University,Xi'an 710071, China}
	
	\author{Tie-Jun Gao}
	\email{tjgao@xidian.edu.cn}
	\affiliation{School of physics, Xidian University,Xi'an 710071, China}
	
	\date{\today}
	
\begin{abstract}
Recently, the worldwide Pulsar  Timing Array (PTA) collaborations detected a stochastic gravitational wave(GW) background in the nanohertz range, which may originate from the early universe's inflationary phase. So in this work, we investigated induce GWs in the T-model inflation with Gauss-Bonnet coupling. Consider the scenario of traversing a domain wall in moduli space, we take the coupling coefficient to be an approximately step function. Within suitable parameter regions, the model exhibits de Sitter fixed points, which allows inflation to undergo an ultra-slow-roll phase, which causes the power spectrum to exhibit a peak. Such a peak can induce nanohertz GWs, which provids an explanation for the PTA observational data. Furthermore, we consider the case of multiple domain wall crossings, and adopting a double-step coupling function. In this case, the resulting GW spectrum has two peaks with frequencies  around \(10^{-8} \,\text{Hz}\) and  \(10^{-2}\,\text{Hz}\), respectively.  Which can be observed by the PTA and the space GW detectors simultaneously.Additionally, the reentry of the power spectrum peaks into the horizon leads to the collapse into primordial black holes (PBHs). We calculate the abundance of PBHs and found that the masses is in the range of \(10^{-14} \sim 10^{-13} M_\odot\) and around \(10^{-2} M_\odot\) , which constitute significant components of the current dark matter.
\end{abstract}
	
\maketitle  
	
\section{introduction}  
Gravitational waves (GWs), first directly detected by the Laser Interferometer Gravitational-Wave Observatory (LIGO) and Virgo Collaborations in 2015, have inaugurated a new era in observational astronomy and cosmology \cite{Abbott_2016}. These ripples in spacetime, as predicted by Einstein’s general relativity, provide a unique tool to study astrophysical phenomena such as black hole and neutron star mergers. Beyond these GWs observed by ground-based interferometers in nanohertz frequency, the PTA, such as NANOGrav \cite{Agazie_2023,Agazi_2023,Afzal_2023}, EPTA \cite{Antoniadis_2023,2023}, CPTA\cite{Xu_2023} and PPTA \cite{zic2023parkespulsartimingarray,Reardon_2023} find the evidence of the stochastic GW background, which offer new opportunities to probe the early universe and large-scale cosmological processes. 

On the other hand, scalar-induced GWs have attracted considerable attention as an intriguing observational target, which are generated through the coupling of large-scale density perturbations during the radiation-dominated era, providing insights into the physics of the primordial Universe. In addition, scalar-induced GWs serve as a complementary probe for understanding the formation of structures such as PBHs \cite{Kohri_2018, Young_2013, PhysRevLett.116.201301, PhysRevLett.117.061101, De_Luca_2021}.Recently, inflationary models with inflection points have been proposed to enhance the power spectrum at small scales \cite{Garc_a_Bellido_2017,Ballesteros_2018,Gao_2018,Xu_2020,Gao_2021}, leading to observable scalar induced GWs, which can also contribute to the formation of PBHs \cite{germani2017primordialblackholesinflection, Di_2018, Garc_a_Bellido_2017}. Some studies suggest that PBHs could constitute all or a significant fraction of the dark matter in the present Universe \cite{PhysRevD.92.023524, PhysRevD.94.083504, PhysRevD.96.043504}. 

To further extend our understanding of early universe, modifications to general relativity have been proposed, including the incorporation of higher-order curvature corrections. Among these, the Gauss-Bonnet (GB) term—a quadratic curvature invariant—has garnered significant attention for its role in string theory \cite{De_Felice_2010}. Incorporating the GB term in inflationary models modifies both the background evolution and the dynamics of perturbations, leading to distinct observational imprints on the stochastic gravitational wave background \cite{Kobayashi_2011}.Recent studies shown that under certain parameter choices in models with GB connection, the cosmological solutions contains nontrivial de sitter fixed point. Similar to the models with inflection points, there is an ultra-slow-roll phase during inflation, which enhances the scalar power spectrum at small scales and induce GWs. Moreover, the GB term has been explored in the context of PBH formation \cite{PhysRevD.104.083545,Kawai_2021} and the generation of scalar-induced GWs in the E-model inflation  with a GB term \cite{PhysRevD.105.063539}. 

In this study, we focus on analyzing the T-model inflation with Gauss-Bonnet coupling. Consider a wall crossing like behavior in the moduli space, the coupling is taken as a Tanh function\cite{Kawai_2021,Antoniadis_1992,Harvey_1996}. For some parameter space, a de Sitter fixed point will appear in the dynamic system, and the corresponding ultra-slow-rolling stage will induce observable GWs. Furthermore, if there are double crossing processes in the moduli space, the coupling will take the form of a double-step function, which will induce GWs with double peaks. We numerically calculate the GW energy spectrum and also calculate the abundance of PBHs produced when the peak re-enter the horizon. In Sec. \ref{1}, we present the T-model inflation with step-like GB coupling functions, and estimate the power spectrum for several parameter sets. In Sec. \ref{2}, we calculate the scalar-induced GWs. In Sec. \ref{3}, we calculate the mass and abundance of PBHs numerically. Finally, we conclude the paper in Sec. \ref{4}.

\section{T-model inflation with Gauss-Bonnet correction}\label{1}
	
	We consider the action of the inflation with the Gauss-Bonnet term is
\begin{equation}
		S =\int   d^4x \sqrt{-g} \left[\frac{1}{2} R - \frac{1}{2}g^{\mu\nu} \partial_\mu \phi \partial_\nu \phi - V(\phi) - \frac{1}{16} \zeta(\phi)  \mathcal{G}_{GB} \right],
\end{equation}
\noindent where the Planck mass \(M_p\) is set to 1, and \(\zeta(\phi)\) is the coupling function, \(\mathcal{G}_{GB} = R^2 - 4R_{\mu\nu}R^{\mu\nu} + R_{\mu\nu\rho\lambda}R^{\mu\nu\rho\lambda} \) is the Gauss-Bonnet term. In this article, we choose the potential to be the T-model attractor inflation
\begin{equation}
	V(\phi) = V_0 \mathrm{Tanh}[m_1 \phi]^{2n}.
\end{equation}
\noindent Here, we take the Gauss-Bennet coupling functions as \cite{Kawai_2021}
\begin{equation}
	\zeta(\phi) = \frac{1}{8} \zeta_0 \mathrm{Tanh}[\zeta_1(\phi-\phi_c)],
\end{equation}
Such step like coupling form is motivated as follows. In calculable cases of type II and heterotic compactifications, the \(\mathcal{G}_{GB}\) coupling to moduli typically arises from one-loop gravitational threshold corrections, determined by the spectrum of BPS states \cite{Antoniadis_1992,Harvey_1996}. When a moduli \(\phi\) crosses the domain wall between different spectra, the coupling function \(\zeta(\phi)\) can approximated by a step like function, for example a Tanh form.
  
In addition, if the moduli space have multiple minimum, then $\Phi$ can cross domain-walls multiple times. For example, we consider a double Tanh function coupling as
\begin{equation}
	\zeta(\phi) = \zeta_{01} \mathrm{Tanh}[\zeta_{11}(\phi-\phi_{c1})] + \zeta_{02} \mathrm{Tanh}[\zeta_{12}(\phi-\phi_{c2})].
\end{equation}
In the context of cosmological models involving a scalar field \(\phi\), the dynamics of the universe can be described by Friedmann euqation and the scalar field evolution equation as

\begin{figure}[h]
	\centering
	\begin{subfigure}{0.48\textwidth}
		\centering
		\includegraphics[width=\textwidth]{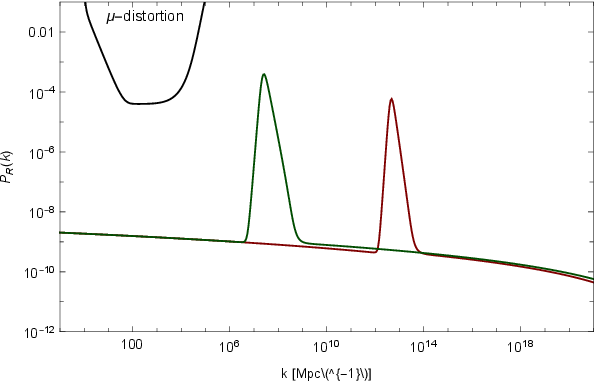}
		\caption{The case of single Tanh coupling with    parameters Set $\mathrm{I}$ (left) and Set $\mathrm{II}$ (right) in TABLE $\mathrm{I}$. }
		\label{fig:11}
	\end{subfigure}
	\hfill
	\begin{subfigure}{0.48\textwidth}
		\centering
		\includegraphics[width=\textwidth]{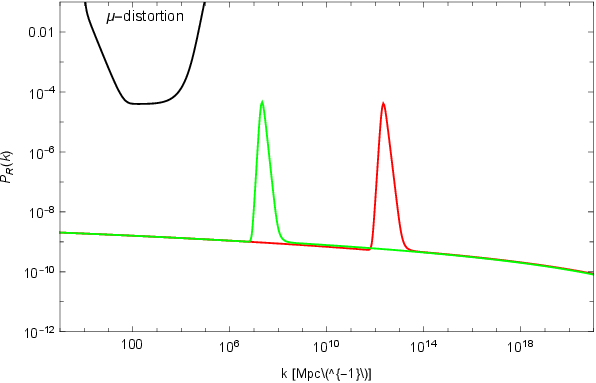}
		\caption{Single Tanh coupling with parameters Set $\mathrm{III}$(left) and Set $\mathrm{IV}$(right) in TABLE $\mathrm{I}$.}
		\label{fig:12}
	\end{subfigure}
	\vspace{0.5cm}
	\begin{subfigure}{0.65\textwidth} 
		\centering
		\includegraphics[width=\textwidth]{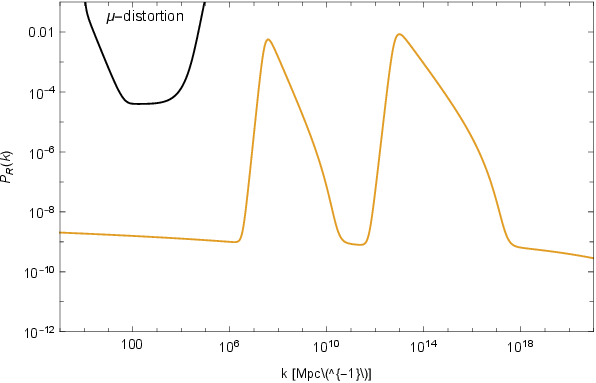}
		\caption{The case of double Tanh coupling with parameters Set $\mathrm{V}$ in TABLE $\mathrm{II}$.}
		\label{fig:13}
	\end{subfigure}
	
	\caption{The power spectrum generated by the single Tanh coupling(1.(a), 1.(b)) and double Tanh coupling(1.(c)). The black curve in the top represents the \(\mu\)-distortion.}
	\label{fig:1}
\end{figure}

\begin{equation}
	3H^2 = \frac{1}{2}\dot{\phi}^2 + V + \frac{3}{2}H^3\zeta_{,\phi}\dot{\phi}, \label{eq3}
\end{equation}

\begin{equation}
	\ddot{\phi} + 3H\dot{\phi} + V_{,\phi} + \frac{3}{2}H^2\left(\dot{H} + H^2\right)\zeta_{,\phi} = 0. \label{eq4}
\end{equation}
\noindent It can be seen that the contribution of the GB term arises entirely from \(\zeta_{,\phi}\). For some parameter space, the dynamic system have de sitter fixed points where \(\dot{\phi} = 0 \),  \(\ddot{\phi} = 0 \), and  \(\dot{H} = 0 \). At the fixed point, the dynamic equations can be simplified as follows

\begin{equation}
	3H^2 = V,
\end{equation}

\begin{equation}
	V_{,\phi} + \frac{3}{2} H^4 \zeta_{,\phi} = 0.
\end{equation}
Then we get that the nontrivial fixed point \(\phi_*\) satisfies
\begin{equation}
(V_{,\phi} + V^2 \zeta_{,\phi}) \Big|_{\phi = \phi_*} = 0.
\end{equation}
Near the fixed point, the inflation will go through an ultra-slow-roll phase. However, to produce reasonable GWs, the fixed point is not an exact one. Due to the presence of the ultra-slow-roll phase, it is always useful to define the Hubble slow-roll parameters \cite{Schwarz_2001}

\begin{table}[t]
	\centering
	\begin{tabular}{|c|c|c|c|c|c|c|c|c|c|c|c|} 
		\hline
		$\mathrm{Sets}$ & $n$ & $V_0/10^{-11}$ & $m_1$ & $\zeta_0/10^7$ & $\zeta_1$ & $\phi_c$ \\
		\hline 
		$\mathrm{I}$ & $0.25$ & $6.91$ & $\sqrt{2/7}$ & $1.25$ & $-200.043$ & $3.52$ \\
		\hline
		$\mathrm{II}$ & $0.25$ & $6.91$ & $\sqrt{2/7}$ & $2.4$ & $-169.806$ & $3.15$  \\
		\hline 
		$\mathrm{III}$ & $1$ & $3.4$ & $\sqrt{1/2}$ & $0.9$ & $-460.972$ & $4.105$   \\
		\hline 
		$\mathrm{IV}$ & $1$ & $3.4$ & $\sqrt{1/2}$ & $1.8$ & $-302.818$ & $3.88$ \\
		\hline
	\end{tabular}
	\caption{The parameters sets for the single Tanh coupling.}
	\label{tab:1}
\end{table}

\begin{table}[t]
	\centering
	\begin{tabular}{|c|c|c|c|c|c|c|c|c|c|c|c|c|c|}
		\hline
		$\mathrm{Sets}$ & $n$ & $V_0/10^{-9}$ & $m_1$ & $\zeta_{01}/10^6$ & $\zeta_{11}$ & $\phi_{c1}$ & $\zeta_{02}/10^6$ & $\zeta_{12}$ & $\phi_{c2}$ \\
		\hline 
		$\mathrm{V}$ & $0.25$ & $6.6$ & $0.01$ & $-2.097$ & $30.3078$ & $4.4$ & $-3.02$ & $23.7829$ & $4$  \\
		\hline
	\end{tabular}
	\caption{The parameters set for the double Tanh coupling.}
	\label{tab:2}
\end{table}

\begin{equation}
\epsilon_1 \equiv -\frac{\dot{H}}{H^2}, \quad \epsilon_{i>1} \equiv \frac{\dot{\epsilon}_{i-1}}{H \epsilon_{i-1}}.
\end{equation}
Moreover, since the presence of the GB term, the additional degrees of freedom suggest the definition of another set of slow-roll parameters \cite{Burgo_2010}	
\begin{equation}
	 \delta_1 \equiv H \dot{\zeta}, \quad \delta_{i>1} \equiv \frac{\dot{\delta}_{i-1}}{H \delta_{i-1}}.
\end{equation}
The Fourier mode \( v_k \) of scalar perturbations satisfies the following equation \cite{Hwang_2000}
\begin{equation}
	v_k'' + \left( c_s^2 k^2 - \frac{z''}{z} \right) v_k = 0.
\end{equation}
where \( c_s^2 \) and \( z^2 \) can be expressed in terms of the slow-roll parameters as
\begin{align}
	z^2 &= a^2 \left( \frac{1 - \delta_1 / 2}{1 - 3 \delta_1 / 4} \right)^2 
	\left( 2 \epsilon_1 - \frac{1}{2} \delta_1 \epsilon_1 + \frac{1}{2} \delta_1 \delta_2 - \frac{1}{2} \delta_1 \epsilon_1 + \frac{3}{4} \frac{\delta_1^2}{2 - \delta_1} \right), \\
	c_s^2 &= 1 - \frac{a^2}{z^2} \left( \frac{\delta_1}{2 - 3 \delta_1 / 2} \right)^2 
	\left( 2 \epsilon_1 + \frac{1}{4} \delta_1 - \frac{1}{4} \delta_1 \delta_2 - \frac{5}{4} \delta_1 \epsilon_1 \right).
\end{align}
\noindent Under the Bunch-Davies vacuum condition
\begin{equation}
		\lim_{\tau \to -\infty} v_k(\tau) = \frac{1}{\sqrt{2c_s k}} e^{-ic_s k \tau}.
\end{equation}
\noindent We finally get the power spectrum through
\begin{equation}
	\mathcal{P_R} = \frac{k^3}{2 \pi^2} \left|\frac{v_k}{8}\right|^2
\end{equation}

\begin{table}[t]
	\centering
	\begin{tabular}{|c|c|c|c|c|c|c|c|c|c|c|c|} 
		\hline
		$\mathrm{Sets}$ &  $n_s$ & $r$ & $\ln(10^{10} A_s)$ & $N^*$ & $M^{peak}_{PBHS}/M_\odot$ & $\Omega_{PBH}/ \Omega_{DM}$\\
		\hline 
		$\mathrm{I}$ &  $0.964$ & $0.002$ & $3.041$ & $63.860$ & $4.9\times10^{-3}$ & $2.11\times10^{-2}$\\
		\hline
		$\mathrm{II}$ &  $0.964$ & $0.002$ & $3.041$ & $62.723$ & $1.1\times10^{-13}$ & $2.84\times10^{-2}$\\
		\hline 
		$\mathrm{III}$ &  $0.967$ & $0.001$ & $3.042$ & $69.125$ & $3.2\times10^{-3}$ & $1.08\times10^{-2}$\\
		\hline 
		$\mathrm{IV}$ & $0.967$ & $0.001$ & $3.042$ & $69.761$ & $4.1\times10^{-13}$ & $5.8\times10^{-2}$\\
		\hline
	\end{tabular}
	\caption{Numerical results obtained from the parameters listed in Table $\mathrm{I}$ for the single  Tanh coupling.}
	\label{tab:3}
\end{table}

\begin{table}[t]
	\centering
	\begin{tabular}{|c|c|c|c|c|c|c|c|c|c|c|c|c|c|}
		\hline
		$\mathrm{Sets}$ & $n_s$ & $r$ & $\ln(10^{10} A_s)$ & $N^*$ & $M^{peak}_{PBHS}/M_\odot$ & $\Omega_{PBH}/ \Omega_{DM}$ & $M^{peak}_{PBHS}/M_\odot$ & $\Omega_{PBH}/ \Omega_{DM}$\\
		\hline 
		$\mathrm{V}$ & $0.966$ & $0.052$ & $3.040$ & $63.398$ &  $2.9\times10^{-14}$ & $0.164$ & $2.1\times10^{-3}$ & $3.1\times10^{-3}$\\
		\hline
	\end{tabular}
	\caption{Numerical results from the parameters listed in Table $\mathrm{II}$ for the double Tanh coupling.}
	\label{tab:4}
\end{table}

Using the parameters Sets in Table $\mathrm{I}$ and Table $\mathrm{II}$, we show the corresponding power spectrum in Fig. 1. (a) and (b) represent the power spectra corresponding to different sets of parameters chosen for the single Tanh coupling, while Fig.1.(c) shows the power spectrum for the double Tanh coupling. It can be seen that the enhancement takes place when the inflation approaches the ultra-slow-roll regime near the de sitter fixed point. As expected, the double Tanh coupling function produces a double peak power spectrum. The numerical results are presented in Tables $\mathrm{III}$ and $\mathrm{IV}$, which are consistent with the latest CMB constraints from Planck 2018, \(n_s = 0.9649 \pm 0.0042\), \(r < 0.064\), and \(\ln(10^{10}A_s) = 3.044 \pm 0.014\) \cite{Dirkes_2018}. 

\section{Gravitational Waves Induced by Scalar Perturbations}\label{2}
	
In this section, we review the GWs induced by scalar perturbations, for more details
in Ref.\cite{Baumann_2007,Ananda_2007,Ando_2018,Espinosa_2018}. First, in the conformal Newtonian gauge, the perturbation metric is expressed as
\begin{equation}
	ds^2 = -a^2(1 + 2\Psi)d\eta^2 + a^2[(1 - 2\Psi)\delta_{ij} + \frac{1}{2}h_{ij}]dx^idx^j,
\end{equation}
where \(\Psi\) represents scalar perturbations and \(h_{ij}\) denotes thesor perturbations. Furthermore, the tensor perturbation can be expanded into plane waves via Fourier transformation
\begin{equation}
	h_{ij}(\eta, \mathbf{x}) = \int \frac{d^3k}{(2\pi)^{3/2}} e^{i\mathbf{k} \cdot \mathbf{x}} \left[h_+(\eta)e_{ij}^+(\mathbf{k}) + h_\times(\eta)e_{ij}^\times(\mathbf{k})\right],
\end{equation}
	where \(h_+(\eta)\) and \(h_\times(\eta)\) are the time-dependent of the two polarization modes, while  \(e_{ij}^+(\mathbf{k})\) and \(e_{ij}^\times(\mathbf{k})\) are the corresponding polarization tensors,expressed as
	\begin{equation}
		e_{ij}^{(+)}(\mathbf{k}) = \frac{1}{\sqrt{2}} \left[ e_i(\mathbf{k}) e_j(\mathbf{k}) - \bar{e}_i(\mathbf{k}) \bar{e}_j(\mathbf{k}) \right],
	\end{equation}
	\begin{equation}
		e_{ij}^{(\times)}(\mathbf{k}) = \frac{1}{\sqrt{2}} \left[ e_i(\mathbf{k}) \bar{e}_j(\mathbf{k}) + \bar{e}_i(\mathbf{k}) e_j(\mathbf{k}) \right].
	\end{equation}
Next, the dynamical equation of tensor modes is obtained from Einstein field equation
	\begin{equation}
		h_k''(\eta) + 2\mathcal{H}h_k'(\eta) + k^2h_k(\eta) = S_k(\eta),
	\end{equation}
	where \(\mathcal{H} = \frac{a'}{a}\) is the conformal Hubble parameter, and	\(S_k(\eta)\) is the source term generated by second-order scalar perturbations, given by
	\begin{equation}
		S_k(\eta) = 4 \int \frac{d^3p}{(2\pi)^{3/2}} e_{ij}(\mathbf{k}) p_i p_j \left[2\Psi_{\mathbf{p}} \Psi_{\mathbf{k} - \mathbf{p}} + \frac{4}{3(1+w)\mathcal{H}^2} (\Psi'_{\mathbf{p}} + \mathcal{H} \Psi_{\mathbf{p}}) (\Psi'_{\mathbf{k} - \mathbf{p}} + \mathcal{H} \Psi_{\mathbf{k} - \mathbf{p}}) \right].
	\end{equation}
	The source term reveals the correlation between GW production and scalar perturbations. Here \(\Psi_{\mathbf{k}}\) can be separated into an initial value \(\psi_{\mathbf{k}}\) and a transfer function \(\Psi(k\eta)\) 
	\begin{equation}
		\Psi_{\mathbf{k}} \equiv \Psi(k\eta) \psi_{\mathbf{k}}.
	\end{equation}
To solve the mode function \(h_k(\eta)\) of GWs,we employ the Green's function method, where the Green's function satisfies
	\begin{equation}
		G_k''(\eta, \eta') + \left(k^2 - \frac{a''(\eta)}{a(\eta)}\right) G_k(\eta, \eta') = \delta(\eta - \eta').
	\end{equation}
	During the radiation-dominated era, the Green's function takes the form
	\begin{equation}
		G_k(\eta, \eta') = \frac{1}{k} \sin[k(\eta - \eta')].
	\end{equation}
Thus, the solution for \(h_k(\eta)\)  can be written as
	\begin{equation}
		h_k(\eta) = \frac{1}{a(\eta)} \int d\eta' G_k(\eta, \eta') a(\eta') S_k(\eta').
	\end{equation}
	This solution fully characterizes the tensor perturbations is induced by scalar perturbation sources.
	Subsequently, we calculate the energy spectrum of GWs. The energy density parameter is expressed as
	\begin{equation}
		\Omega_{\text{GW}}(\eta, k) = \frac{1}{24} \left(\frac{k}{\mathcal{H}}\right)^2 \overline{\mathcal{P}_h(\eta, k)},
	\end{equation}
	where \(\overline{\mathcal{P}_h(\eta, k)}\) represents the oscillation average of tensor power spectrum.Assuming \(\psi_{\mathbf{k}}\) satisfy Gaussian distribution,and defining three dimensionless variables 
	\( u \equiv |\mathbf{k} - \mathbf{p}|/k \), \( v \equiv |\mathbf{p}|/k \) ,and \( x \equiv k\eta \) , then the power spectrum can be expressed as 
	\begin{equation}
		\mathcal{P}_h(\eta, k) = 4 \int_{0}^{\infty} dv \int_{|1-v|}^{1+v} du \left( \frac{4v^2 - (1 + v^2 - u^2)^2}{4uv} \right)^2 \mathcal{I}^2(x, u, v) \mathcal{P}_R(ku) \mathcal{P}_R(kv). 
	\end{equation} 
	In the limit \( x \to \infty \) , \(\mathcal{I}(x, u, v) \) in the radiation dominated era becomes
	\begin{multline}
		\mathcal{I}_{RD}(x \to \infty, u, v) = \frac{3 \left( u^2 + v^2 - 3 \right)}{4u^3v^3x} \left\{ \sin x \left[ -4uv + \left( u^2 + v^2 - 3 \right) \log \left| \frac{3 - (u + v)^2}{3 - (u - v)^2} \right| \right] \right.\\ 
		\left. - \pi \left( u^2 + v^2 - 3 \right) \Theta(v + u - \sqrt{3}) \cos x \right\}
	\end{multline}
After oscillation averaging
	\begin{multline} 
		\overline{\mathcal{I}^2_{RD}(x \to \infty, u, v)} = \frac{1}{2} \left( \frac{3}{4u^3v^3x} \right)^2  \\
		(u^2 + v^2 - 3)^2 \left\{ \left[ -4uv + (u^2 + v^2 - 3) \ln \left| \frac{3 - (u + v)^2}{3 - (u - v)^2} \right| \right]^2 + \left[ \pi (u^2 + v^2 - 3) \Theta(u + v - \sqrt{3}) \right]^2 \right\}. 
	\end{multline}  
	Combining Eqs. (27) and (28), with \(\mathcal{H} = 1/\eta\) in the radiation-dominated era,the final expression of the GW energy spectrum is 
	\begin{multline} 
		\Omega_{GW}(\eta, k) = \frac{1}{12} \int_0^{\infty} dv \int_{|1-v|}^{1+v} du \left( \frac{4v^2 - (1 + v^2 - u^2)^2}{4uv} \right)^2 \\
		\mathcal{P}_R(ku) \mathcal{P}_R(kv) \left( \frac{3}{4u^3v^3} \right)^2 (u^2 + v^2 - 3)^2 \\
		\left\{ \left[ -4uv + (u^2 + v^2 - 3) \ln \left| \frac{3 - (u + v)^2}{3 - (u - v)^2} \right| \right]^2 \right. \\
		+ \left. \left[ \pi (u^2 + v^2 - 3) \Theta(u + v - \sqrt{3}) \right]^2 \right\}.
	\end{multline}

To study the energy spectrum of induced GWs, we connect the comoving wave number \(k\) to the present-day frequency \(f\), which is given by
\begin{equation}
	f \approx 0.03~\mathrm{Hz} \cdot \frac{k}{2 \times 10^7~\mathrm{pc}^{-1}}.
	\label{eq:frequency_relation}
\end{equation}
And the GWs energy density parameter at the present time, \(\Omega_{\mathrm{GW},0}\) can be calulated by
\begin{equation}
	\Omega_{\mathrm{GW},0} = 0.83 \left(\frac{g_{*,0}}{g_{*,p}}\right)^{-1/3} \Omega_{r,0} \Omega_{\mathrm{GW}},
	\label{eq:omega_gw_present}
\end{equation}

\begin{figure}[h]
	\centering
	\begin{subfigure}{0.48\textwidth}
		\centering
		\includegraphics[width=\textwidth]{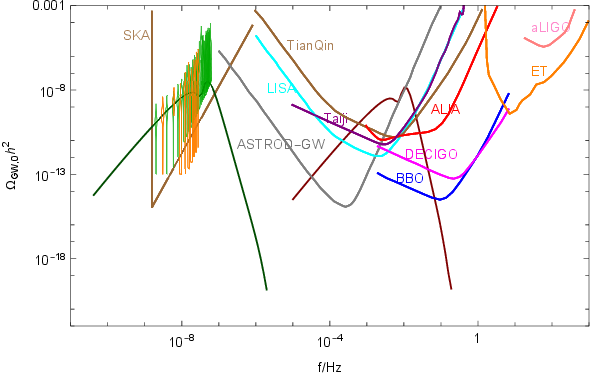}
		\caption{The GW spectrum from the single Tanh coupling of Set $\mathrm{I}$(left) and Set $\mathrm{II}$(right).}
		\label{fig:21}
	\end{subfigure}
	\hfill
	\begin{subfigure}{0.48\textwidth}
		\centering
		\includegraphics[width=\textwidth]{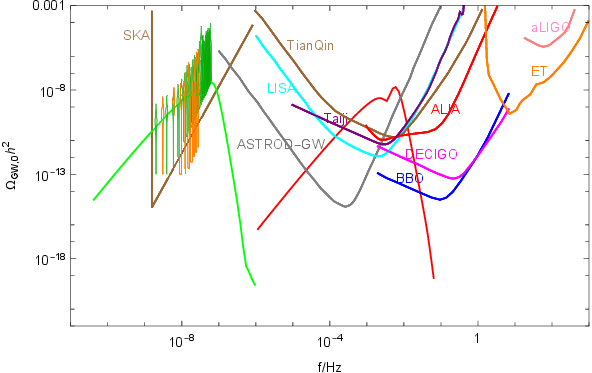}
		\caption{The GW spectrum from the single Tanh coupling of Set $\mathrm{III}$(left) and Set $\mathrm{IV}$(right).}
		\label{fig:22}
	\end{subfigure}
	\vspace{0.5cm}
	\begin{subfigure}{0.65\textwidth} 
		\centering
		\includegraphics[width=\textwidth]{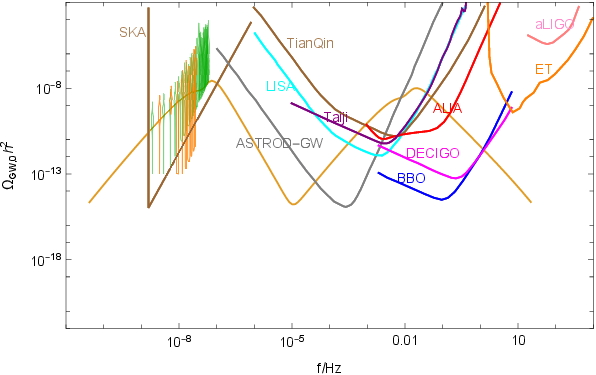}
		\caption{The double peak GW spectrum obtained from the double Tanh coupling corresponding to Set $\mathrm{V}$.}
		\label{fig:23}
	\end{subfigure}
	\caption{The energy spectrum of the induced GWs predicted by the single Tanh coupling(2(a) and 2(b))and the double Tanh coupling (2(c)).The expected sensitivity curves above from  Ref.\cite{Moore_2014,McLaughlin_2013,Hobbs_2013,Luo_2016,amaroseoane2017laserinterferometerspaceantenna}. The green region represents the constraint from NANOGrav\cite{Arzoumanian_2020}, while the orange region corresponds to the constraint from EPTA\cite{Ferdman_2010}}
	\label{fig:2}
\end{figure}

\noindent where \(\Omega_{r,0} \simeq 9.1 \times 10^{-5}\) is the current density fraction of radiation, \(g_{*,0}\) and \(g_{*,p}\) are the effective degrees of freedom for energy density at the present time and at the time when the peak  mode crosses the horizon, respectively.

In Fig.2, we present the numerical results of the GW spectrum for different cases. Recently, the PTA observed stochastic GW background around nanohertz , and the corresponding observational data represented in the green and orange regions in the figure. It can be seen that in Fig.2(a) and (b), the lower ferquency peak on the left (approximately \(10^{-8} - 10^{-7}\,\text{Hz}\)) is from the single Tanh coupling with parameter Set$\mathrm{I}$ and $\mathrm{III}$, which is consistent with the PTA data.And the higher frequency peak on the right (around \(10^{-2}\,\text{Hz}\)) is from parameter set$\mathrm{II}$ and $\mathrm{IV}$, they are both above the expected sensitivity curves of LISA, Taiji, and TianQin etc. In addtion, as show in Fig.2.(c), for the double Tanh coupling ,the GWs spectrum have two peaks. If the GWs corresponding to the lower-frequency peak are detected by PTA,the higher frequency peak should be detected by other detectors such as LISA. Therefore the double peaks model can be distinguished from other single peak models. 

\section{Primordial Black Hole Formation}  \label{3}

If the amplitude of small-scale primordial density perturbations is sufficiently large, when these perturbations re-entering the Hubble horizon during the radiation-dominated era, it will undergo gravitational collapse, and leading to the formation of PBHs, which provides an explanation for the origin of dark matter\cite{GreenKavanagh2021,Sasaki2018,Hertzberg_2018}.  

\begin{figure}[t]
	\centering
	\begin{subfigure}{0.48\textwidth}
		\centering
		\includegraphics[width=\textwidth]{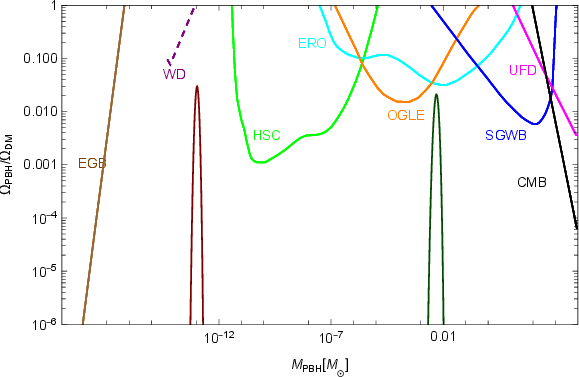}
		\caption{ The PBH abundance corresponding to the single Tanh coupling function for Set $\mathrm{I}$(right) and Set $\mathrm{II}$(left) of Table $\mathrm{I}$.}
		\label{fig:31}
	\end{subfigure}
	\hfill
	\begin{subfigure}{0.48\textwidth}
		\centering
		\includegraphics[width=\textwidth]{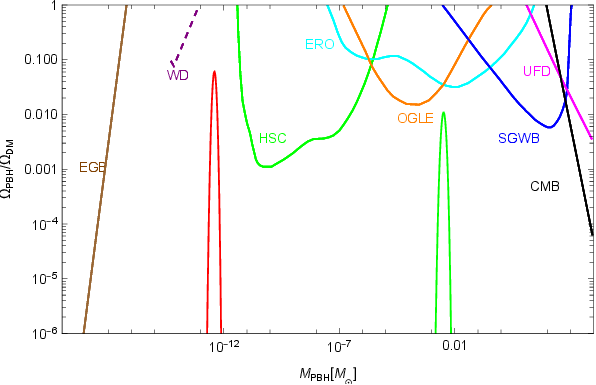}
		\caption{ The PBH abundance corresponding to the single Tanh coupling function for Set $\mathrm{III}$(right) and Set $\mathrm{IV}$(left) of Table $\mathrm{I}$.}
		\label{fig:32}
	\end{subfigure}
	\vspace{0.5cm}
	\begin{subfigure}{0.65\textwidth} 
		\centering
		\includegraphics[width=\textwidth]{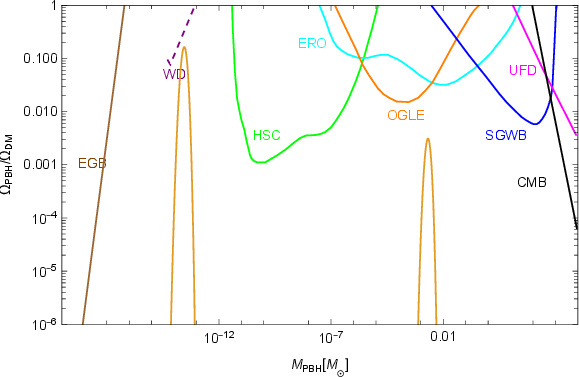}
		\caption{The PBH abundance corresponding to the double Tanh coupling function for Set $\mathrm{V}$ of Table $\mathrm{II}$.}
		\label{fig:33}
	\end{subfigure}
	
	\caption{The numerical results of PBH abundance $\Omega_{PBH}/ \Omega_{DM}$ for the single Tanh coupling are shown in 3.(a) and 3.(b), while the result for the double Tanh coupling is presented in 3.(c).The constraint data are obtained from references \cite{Green_2021,PhysRevD.100.063521,Niikura_2019,PhysRevLett.123.251101,PhysRevLett.125.101101,PhysRevLett.111.181302,PhysRevD.96.123523,PhysRevD.95.043534}.}
	\label{fig:3}
\end{figure}

The mass of a PBH is proportional to the horizon mass \(M_H\) at the time when the perturbation re-enters the Hubble horizon, expressed as \(M = \gamma M_H\), where \(\gamma \approx 0.2\) is the collapse efficiency factor\cite{1975ApJ...201....1C}, representing the fraction of perturbation mass that forms black holes. The horizon mass depends on the radiation energy density \(\rho\) and the Hubble parameter \(H\)  
\begin{equation}  
	M_H = \frac{4}{3} \pi \rho H^{-3}.  
\end{equation}  
This relationship can be approximated in terms of the wavenumber \(k\) as \cite{Ballesteros_2018}  
\begin{equation}  
	M \simeq 10^{18} \, \text{g} \left( \frac{\gamma}{0.2} \right) \left( \frac{g_*}{106.75} \right)^{-1/6} \left( \frac{k}{7 \times 10^{13} \, \text{Mpc}^{-1}} \right)^{-2},  
\end{equation}  
where \(g_*\) represents the effective degrees of freedom.

Assuming that the density perturbations follow a Gaussian distribution, the PBH formation rate \(\beta(M)\) can be expressed in terms of the standard deviation \(\sigma(M)\) of the perturbations 
\begin{equation}  
	\beta(M) = \frac{1}{\sqrt{2\pi \sigma^2(M)}} \int_{\delta_c}^{\infty} \exp\left(-\frac{\delta^2}{2\sigma^2(M)}\right) d\delta,  
\end{equation}  
where \(\delta_c \approx 0.45\) is the critical collapse threshold \cite{PhysRevD.88.084051} and \(\sigma^2(M)\) is computed from the primordial curvature perturbation power spectrum \(\mathcal{P}_R\) in Sec \(\mathrm{II}\)   
\begin{equation}  
	\sigma^2(M) = \frac{16}{81} \int \frac{dq}{q} (qR)^4 P_R(q) W^2(qR),  
\end{equation}  
the Gaussian smoothing window function is \(W(x) = exp(-x^2/2)\).
Using the error function, the formation rate can be simplified as  
\begin{equation}  
	\beta(M) = \frac{1}{2} \, \text{erfc} \left( \frac{\delta_c}{\sqrt{2} \sigma(M)} \right). 
\end{equation}  

The current abundance of PBHs in the universe, \(\Omega_{\text{PBH}}\), can be obtained by integrating the formation rate over all mass ranges  
\begin{equation}  
	\Omega_{\text{PBH}} = \int \frac{dM}{M} \Omega_{\text{PBH}}(M). 
\end{equation}  
The expression for its relative abundance to dark matter is given by
\begin{equation}  
	\frac{\Omega_{\text{PBH}}(M)}{\Omega_{\text{DM}}} = \left( \frac{\beta(M)}{1.6 \times 10^{-16}} \right) \left( \frac{\gamma}{0.2} \right)^{3/2} \left( \frac{g_*}{106.75} \right)^{-1/4} \left( \frac{M}{10^{18} \, \text{g}} \right)^{-1/2},  
\end{equation}  
where the total abundance \(\Omega_{\text{DM}} \approx 0.26\) \cite{Musco_2013} .The results of the numerical calculations are presented in Fig. 3, Table $\mathrm{III}$ and Table $\mathrm{IV}$. From Fig.3,it can be seen that when the coupling is a single Tanh function in Eq(3), the peak masses of PBH for parameters Sets\(\mathrm{I}\) and \(\mathrm{II}\) are $4.9 \times 10^{-3} M_\odot$ and $1.1 \times 10^{-13} M_\odot$, with corresponding abundances are $2.11 \times 10^{-2}$ and $2.84 \times 10^{-2}$, respectively. For parameter sets \(\mathrm{III}\) and \(\mathrm{IV}\), the peak masses are $3.2 \times 10^{-3} M_\odot$ and $4.1 \times 10^{-13} M_\odot$, with corresponding abundances are $1.08 \times 10^{-2}$ and $5.8 \times 10^{-2}$. When the coupling function is the double tanh function in Eq(4), as shown in the Fig.3.(c), it can produces PBHs with two different masses simultaneously, with the peak masses are $2.9 \times 10^{-14} M_\odot$ and $2.1 \times 10^{-3} M_\odot$, corresponding abundances are 0.164 and $3.1 \times 10^{-3}$, respectively. Therefore, the results show that the PBHs can constitute significant components of the current dark matter.

\section{summary}\label{4}

In this study, we introduce the coupling between the scalar field and the Gauss-Bonnet term in the T-model inflation. Consider the coupling has the form of a single Tanh and double Tanh functions respectively, the inflation will undergo an ultra-slow-roll phase which results in the peaks of power spectrum, which subsequently induces GWs with single peak or double peaks. These enhanced of  also lead to the gravitational collapse and forming PBHs when reentry the horizon.

Numerically, we computed the energy spectrum of GWs and demonstrated that for parameter Sets \(\mathrm{I}\) and \(\mathrm{III}\) of the single peak model, as well as for Set \(\mathrm{V}\) of the double peak model, the peaks fall within the region of the PTA data.Therefore, these scalar induced gravitational waves can be regarded as the source of the PTA signal. Meanwhile, for Set \(\mathrm{II}\) and \(\mathrm{IV}\) of single peak model and Set \(\mathrm{V}\) of double peak model,the peak above the anticipated sensitivity curves of LISA, Taiji, and TianQin etc., which can be detectable in the near future. And the double peak model can be distinguished from other single peak models.

Additionally, we applied the Press-Schechter formalism for gravitational collapse to calculate the abundance of PBHs produced in both models. The results show that for the double Tanh coupling, the peak masses are \(2.9 \times 10^{-14} M_\odot\) and \(2.1 \times 10^{-3} M_\odot\), which can play a substantial role in constituting dark matter.

\begin{acknowledgments}
	
	This work was supported by “the Natural Science Basic Research Program ofShaanxi Province” No. 2023-JC-YB-072. 
	
\end{acknowledgments}

\bibliographystyle{apsrev4-1}
\bibliography{reference.bib}

\end{document}